\documentclass[manuscript]{revtex4}
\usepackage{graphicx}
\usepackage{amsmath}
\usepackage{amssymb}

\newcommand{\ket}[1]{|#1\rangle}
\newcommand{\bra}[1]{\langle #1|}
\newcommand{\braket}[2]{\langle #1|#2\rangle}
\newcommand{\ketbra}[2]{|#1\rangle\langle #2|}

\begin{document}
\title{Quasi-deterministic generation of entangled atoms in a cavity}
\author{Jongcheol Hong and Hai-Woong Lee}
\affiliation{Department of Physics,
         Korea Advanced Institute of Science and Technology,\\
         373-1 Kusong-dong Yusong-gu Taejon 305-701, Korea}
\date{\today}

\begin{abstract}
We present a scheme to generate a maximally entangled state of two
three-level atoms in a cavity. The success or failure of the generation
of the desired entangled state can be determined by detecting the
polarization of the photon leaking out of the cavity. With the use of
an automatic feedback, the success probability of the scheme can be
made to approach unity.
\end{abstract}

\maketitle

There has recently been much interest in the generation of
entangled states of two or more particles, as they give rise to
quantum phenomena that cannot be explained in classical terms.
Entangled states not only are used to test fundamental
quantum-mechanical principles such as Bell's inequalities\cite{1}
but also play a central role in practical applications of the
quantum information theory such as quantum computation\cite{2},
quantum teleportation\cite{3}, quantum dense coding\cite{4} and
quantum cryptography\cite{5}. Two-photon entangled states can
commonly be produced from a nonlinear optical process such as
parametric downconversion\cite{6}. Although there seems no easy
way of generating entangled states of massive particles instead of
massless photons, recent advances in ion trapping technology and
cavity QED have led to several proposals\cite{7} for generation of
entangled states of atoms or ions and subsequent experimental
realizations\cite{8}.

In this paper we introduce a scheme that allows the generation of
a maximally entangled state of two $\Lambda$-type three-level
atoms in a cavity. The scheme is similar to that proposed recently
by Plenio et al.\cite{9}, but has an advantage in that the
probability of obtaining the entangled state can be made to
approach unity, as described below.

\begin{figure}[p]
\includegraphics[scale=0.5]{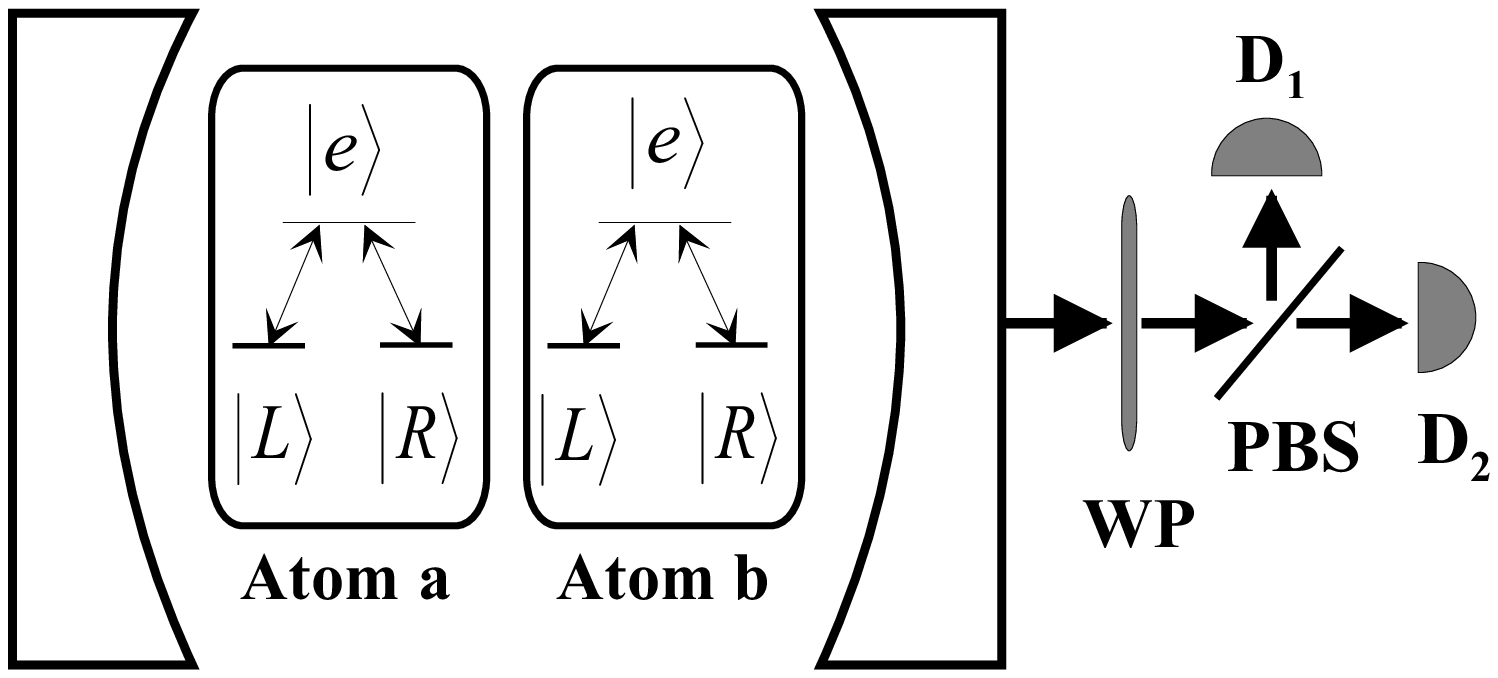}\caption{Experimental
  scheme. WP represents a quater wave plate, PBS denotes a polarization
  beam splitter and $D_1$ and $D_2$ are detectors. If a photon incidents
  on the quarter wave plate is left(right)-circular polarized,
  then it is detected by $D_1$($D_2$)}\label{cavity}
\end{figure}

The system we consider is shown schematically in Fig.\ref{cavity}.
Two identical $\Lambda$-type three-level atoms $a$ and $b$, each
with an excited state $\ket{e}$ and two degenerate ground states
$\ket{L}$ and $\ket{R}$, are situated in a resonant optical
cavity. The $\ket{e} \leftrightarrow \ket{L}$ transition is
coupled by left-circularly polarized light, while the $\ket{e}
\leftrightarrow \ket{R}$ transition by right-circularly polarized
light. We assume that the separation between the two atoms is
large compared with the wavelength of the $\ket{e}\leftrightarrow
\ket{L}$ or $\ket{e}\leftrightarrow \ket{R}$ transition so that
the dipole-dipole interaction can be neglected.

The outline of our scheme is as follows. Initially we prepare the
two atoms in their ``left" ground state $\ket{L}$ and inject a
left-circularly polarized photon into the cavity. One of the two
atoms can then absorb the photon and make an upward transition to
$\ket{e}$. It can subsequently deexcite to $\ket{L}$ or $\ket{R}$
emitting a left- or right-circularly polarized photon. If the
polarization of the emitted photon can be detected and is found
right-circularly polarized, one can conclude that one of the two
atoms is in $\ket{L}$ and the other is in $\ket{R}$. Since which
atom is in $\ket{L}$ and which in $\ket{R}$ cannot be determined,
the final state of the two atoms is a superposition of the two
probabilities, i.e. an entangled state. Thus an entangled state of
the two atoms results when the polarization of the photon leaking
out of the cavity is detected to be right-circularly polarized.

In order to illustrate the main idea, let us first consider an
ideal case of the perfect cavity. The temporal evolution of the
system in the cavity is governed by the Hamiltonian $H = H_a + H_b
+ H_R + H_I$, where $H_a$ and $H_b$ are the atomic Hamiltonian for
atoms $a$ and $b$, respectively, $H_R$ is the free field
Hamiltonian and $H_I$ is the interaction Hamiltonian given by
\begin{equation}\label{eq1}
H_I=i\hbar\sum_{i=a,b}\sum_{\lambda=L,R}(g_\lambda
c_\lambda\ket{e}_{ii}\bra{\lambda}-g_\lambda
c^\dag_\lambda\ket{\lambda}_{ii}\bra{e})
\end{equation}
In Eq.(\ref{eq1}) $c_\lambda$ and $c_\lambda^\dag$ $(\lambda=L,R)$
denote the annihilation and creation operators for the left or
right circularly polarized cavity field,
$g_\lambda~(\lambda=L,R)$, assumed to be real, represents the
coupling strength between the atom and the left or right
circularly polarized field ($g_\lambda$ is assumed to be the same
for atom $a$ and atom $b$), $\ket{e}_i (i=a,b)$ represents the
excited state of the atom $a$ or $b$, and $\ket{\lambda}_i
(\lambda=L,R; i=a,b)$ represents the ``left'' or ``right'' ground
state of the atom $a$ or $b$. Expressing the state of the total
system in the form $\ket{atom~a,~atom~b;~photon}$, the initial
state can be written $\ket{L,L;L}$. Under the rotating wave
approximation, the temporal evolution of the system is spanned by
the five basis states, $\{$ $\ket{L,L;L}$, $\ket{e,L;0}$,
$\ket{L,e;0}$, $\ket{R,L;R}$, $\ket{L,R;R}$ $\}$. A
straightforward algebra yields that the state of the system at
time $t$ is given in terms of these basis states as
\begin{equation}\label{eq2}
\begin{split}
\ket{\Psi(t)}=&\frac{g_R^2+2g_L^2\cos{\alpha t}}{\alpha^2}\ket{L,L;L}\\
&-\frac{g_L}{\alpha}\sin{\alpha t}(\ket{e,L;0}+\ket{L,e;0}) \\
&\mbox{}-\frac{2g_Lg_R}{\alpha^2}\sin^2{\frac{\alpha
t}{2}}(\ket{R,L;R}+\ket{L,R;R})
\end{split}
\end{equation}
Eq.(\ref{eq2}) indicates that the probability at time $t$ of
obtaining the entangled state$\ket{\phi}=\frac{1}{\sqrt{2}}
(\ket{R,L;R}+\ket{L,R;R})$ is given by
\begin{equation}
\begin{split}
P(t)=|\braket{\phi}{\Psi(t)}|^2&=8\frac{g_L^2g_R^2}{\alpha^4}\sin^4\frac{\alpha
  t}{2}\\&=\frac{8\beta^2}{(\beta^2+2)^2}\sin^4\frac{\alpha t}{2}
\end{split}
\end{equation}
where $\beta\equiv g_R/g_L$. At
$t=\frac{(2n+1)\pi}{\alpha}~(n=0,1,2,\cdots)$, the probability has the
maximum value $P_{max}=\frac{8\beta^2}{(\beta^2+2)^2}$. When
  $\beta=1$, $P_{max}=\frac{8}{9}$. In particular, when
  $\beta=\sqrt{2}$, $P_{max}$ reaches $1$.

We now analyze the system depicted in Fig.\ref{cavity}, in which the
polarization of the photon leaking out of the cavity is monitored. In
order to describe the temporal evolution of the open system under
consideration, we adopt the master equation approach. The master
equation describing the time evolution of the density matrix is given
by
\begin{equation}
\frac{d\rho}{dt}=\frac{1}{i\hbar}[H,\rho]
+\frac{\kappa}{2}\sum_{\lambda=L,R}(2c_\lambda\rho
c_\lambda^\dag-c_\lambda^\dag c_\lambda\rho-\rho c_\lambda^\dag
c_\lambda)
\end{equation}
where $\kappa$ denotes the cavity decay rate. Assuming that the
initial state is $\ket{L,L;L}$, the time evolution of the system
inside the cavity is now described with eight basis states, $\{$
$\ket{L,L;L}$, $\ket{e,L;0}$, $\ket{L,e;0}$, $\ket{R,L;R}$,
$\ket{L,R;R}$, $\ket{L,L;0}$, $\ket{R,L;0}$, $\ket{L,R;0}$ $\}$.
Compared with the perfect cavity case previously considered, we
now have three more basis states $\ket{L,L;0}$, $\ket{R,L;0}$ and
$\ket{L,R;0}$ which result when the photon in the states
$\ket{L,L;L}$, $\ket{R,L;R}$ and $\ket{L,R;R}$, respectively,
escapes the cavity. When the detector in Fig.\ref{cavity}
registers a left circularly polarized photon i.e., when $D_1$
clicks, we know for sure that the state of the system in the
cavity is $\ket{L,L;0}$. On the other hand, if a right circularly
polarized photon is detected i.e. when $D_2$ clicks, the state of
the system in the cavity can be $\ket{R,L;0}$ or $\ket{L,R;0}$.
Whether the state is $\ket{R,L;0}$ or $\ket{L,R;0}$ cannot be
determined from the measurement of the polarization of the photon.
Since both the initial state and the system Hamiltonian are
symmetric with respect to the two atoms $a$ and $b$, the state
associated with the detection of right circularly polarized photon
must be $\frac{1}{\sqrt{2}}(\ket{R,L;0}+\ket{L,R;0})$. It is then
clear that, at large time $t\rightarrow\infty$, the system inside
the cavity approaches a mixture of $\ket{L,L;0}$ and
$\frac{1}{\sqrt{2}}(\ket{R,L;0}+\ket{L,R;0})$, i.e.,
\begin{equation}
\begin{split}
\rho_\infty=&(1-p)\ketbra{L,L;0}{L,L;0}\\
&+\frac{p}{2}\left(\ket{R,L;0}+\ket{L,R;0}\right)
\otimes\left(\bra{R,L;0}+\bra{L,R;0}\right)
\end{split}
\end{equation}
where $p$ represents the probability to obtain the desired entangled
state $\frac{1}{\sqrt{2}}(\ket{R,L}+\ket{L,R})$ of the two atoms $a$
and $b$.

\begin{figure}[p]
\includegraphics[scale=0.5]{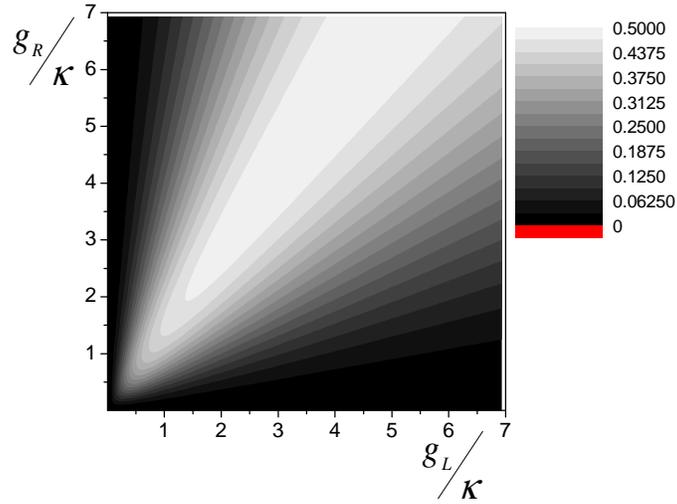}\caption{Probability to
  obtain the entangled state as a function of $g_R/\kappa$ and
  $g_L/\kappa$}\label{prob}
\end{figure}

The probability $p$ is determined when the system parameters
$g_L$, $g_R$ and $\kappa$ are given. In Fig.\ref{prob} we plot $p$
as a function of $g_L/\kappa$ and $g_R/\kappa$ computed from
numerical simulation of the master equation. The probability is
seen to have its maximum value of $\sim\frac{1}{2}$ along the line
$\frac{g_R}{g_L}=\sqrt{2}$.

Our scheme of Fig.\ref{cavity} offers a way of obtaining the
two-atom entangled state $\frac{1}{\sqrt{2}}(\ket{R,L}+\ket{L,R})$
with probability $p$. Our scheme is thus probabilistic. When the
scheme fails to generate the desired entangled state, however, the
experiment can easily be repeated for another round of trial. The
probability with which the scheme fails to generate the entangled
state is $1-p$. In this case the detector registers a
left-circularly polarized photon and the state inside the cavity
is $\ket{L,L;0}$. The experiment can then simply be repeated by
injecting another left-circularly polarized photon into the
cavity. The state of the system inside the cavity is then
$\ket{L,L;L}$ and the entire experiment restarts. In fact, one can
have the experiment automatically repeat itself in case of the
failure simply by eliminating the detector $D_1$ and replacing it
by a path directed back to the cavity, so that the left-circularly
polarized photon can be automatically fed back to the cavity. One
then needs only to wait for the detector $D_2$ to click. The
moment the detector $D_2$ registers a photon, we know that the
entangled state $\frac{1}{\sqrt{2}}(\ket{R,L}+\ket{L,R})$ of the
two atoms $a$ and $b$ is generated in the cavity. The probability
that the entangled state is not generated after $n$ rounds of
trial is $(1-p)^n$. Since the failure probability exponentially
decreases with respect to the number of rounds, the desired
entangled state can be generated with high probability within a
few cavity decay times. We also note that the generated entangled
state is a superposition of different combinations of two ground
states (or metastable states) and thus is free from decoherence
caused by the cavity loss as well as spontaneous emission.

It should be pointed out that our treatment assumes that the only
losses in the system are those associated with the coupling of the
cavity field mode to the outer field modes. Thus, only the losses
of this type lead to photon detection. Losses due to spontaneous
emission into modes other than the cavity mode are neglected.
Absorption of photons by the cavity mirrors are also neglected.

\begin{figure}[p]
\includegraphics[scale=0.5]{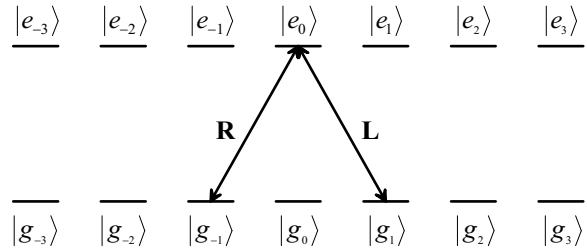}\caption{$(F=3)$ hyperfine
levels of cesium. $R$ and $L$ indicate that the two levels
connected by the arrow are coupled by right- and left- circularly
polarized light, respectively}\label{levels}
\end{figure}

As a specific example for realization of our scheme proposed here,
we consider hyperfine levels of cesium (nuclear spin
$I=\frac{7}{2}$) considered recently by Lange and Kimble\cite{10}.
The Zeeman sublevels of the states ($6S_{1/2},F=3$) and
($6P_{1/2},F=3$) are drawn in Fig.\ref{levels}, where
$\ket{g_{m_F}}$ and $\ket{e_{m_F}}$ denote the sublevels
($6S_{1/2},F=3,m_F$) and ($6P_{1/2},F=3,m_F$), respectively, with
$m_F$ running from $-3$ through $3$. The wavelength of the
$\ket{e}\leftrightarrow\ket{g}$ transition is $852.36nm$. The
transition between $\ket{e_{m_F}}$ and $\ket{g_{m_F-1}}$ is
mediated by right-circularly polarized light and that between
$\ket{e_{m_F}}$ and $\ket{g_{m_F+1}}$ by left-circularly polarized
light. With the cesium atoms prepared initially in
$\ket{g_{m_F+1}}$ and one left-circularly polarized photon
injected into the cavity, one can then obtain, using our scheme,
the entangled state $\frac{1}{\sqrt{2}}(\ket{g_{m_F-1},g_{m_F+1}}+
\ket{g_{m_F+1},g_{m_F-1}})$ of the cesium atom with probability
$p$.

Let us estimate the probability $p$ for the above system. The
Hamiltonian representing the interaction of the cesium atom and
the injected photon can be written as \cite{10}
\begin{equation}\label{eq9}
H_I=-i\hbar g_0\sum_{i=a,b}\sum_{\lambda=L,R}(c_\lambda^\dag
A_{i,\lambda}-A_{i,\lambda}^\dag c_\lambda)
\end{equation}
where $2g_0/2\pi$ is the single-photon Rabi frequency and
$A_{i,\lambda}$ is given by
\begin{equation}\label{eq10}
A_{i,\lambda}=\sum_{m_g,m_e}\ket{F,m_g}_{i~i}
\braket{F,m_g;\lambda}{F,m_e}_{i~i}\bra{F,m_e}
\end{equation}
where $m_g$ and $m_e$ denote the Zeeman sublevel quantum number
$m_F$ for the states $\ket{g_{m_F}}$ and  $\ket{e_{m_F}}$,
respectively. Comparing Eqs.(\ref{eq9}) and (\ref{eq10}) with
Eq.(\ref{eq1}), we find that $g_L$ and $g_R$ of Eq.(\ref{eq1})
correspond to $g_0$ of Eq.(\ref{eq9}) multiplied by the
Clebsch-Gordan coefficient $\braket{F_g,m_g;\lambda}{F_e,m_e}$.
With the present technology, one can achieve the condition
$g_0\cong3\kappa\sim15\kappa$\cite{11} (For example, the values of
$g_0/2\pi\cong 120MHz$ and $\kappa/2\pi\cong 40MHz$ were cited
in\cite{11}). Taking $g_0=3\kappa$ and $\ket{g_1}$ as the initial
atomic state, we obtain $g_L=g_R=\frac{1}{\sqrt{2}}g_0$ and
$p\cong 0.43$. In this case the entangled state
$\frac{1}{\sqrt{2}}(\ket{g_{-1},g_{1}}+ \ket{g_{1},g_{-1}})$ is
obtained with the probability $0.43$ after one round of trial.
Note that in this case there is no chance for the states other
than the states $\ket{e_0}$, $\ket{g_{-1}}$ and $\ket{g_1}$ to be
occupied, because there is one and only one photon present in the
cavity initially (see Fig.\ref{levels}). Taking $g_0=3\kappa$ and
$\ket{g_0}$ as the initial atomic state, we obtain
$g_L=\sqrt{\frac{5}{12}}g_0$, $g_R=\frac{1}{\sqrt{2}}g_0$ and
$p\cong 0.45$. The atomic entangled state obtained in this case is
$\frac{1}{\sqrt{2}}(\ket{g_{-2},g_{0}}+ \ket{g_{0},g_{-2}})$. We
can thus conclude that our scheme with the help of the present
cavity technology allows the generation of the entangled atomic
state with a reasonably high probability even after one round of
trial.

A comparison of our scheme with the scheme proposed by Plenio et
al.\cite{9} is now in order. The scheme of Plenio et al. provides
a way of generating the entangled state
$\frac{1}{\sqrt{2}}(\ket{e,g}-\ket{g,e})$ of two two-level atoms
inside a cavity, where $\ket{e}$ and $\ket{g}$ refer to the upper
and lower levels. Since the generated entangled state is an
antisymmetric trapped state, it is important to prepare the
initial state in a nonsymmetric way, e.g., the initial state can
be $\ket{e,g;0}$, the atom $a$ in $\ket{e}$, the atom $b$ in
$\ket{g}$ and no photon. The success of the scheme depends upon
detection of no photon leaking out of the cavity. If a photon
leaking out of the cavity is detected, then the experiment fails.
In this case the state of the system inside the cavity is
$\ket{g,g;0}$. If another photon is injected into the cavity, then
the state of the system becomes $\ket{g,g;1}$. This state is
symmetric with respect to an interchange of the two atoms. It is
thus clear that the experiment cannot be repeated for another
round of trial by simply reinjecting another photon. Our scheme in
contrast is designed in such a way that the ``correct'' initial
state is set up in case of the failure simply by reinjecting the
photon leaked out of the cavity back into the cavity. Although
probabilistic in nature, our scheme thus provides a
quasi-deterministic way of generating an entangled state of two
atoms.

The advantage of our scheme still prevails even if we take into
account the finite detection efficiency of the detectors. When no
photon is detected in the scheme of Plenio et al., there are two
possibilities : (1) the experiment has succeeded and the desired
entangled state has been generated, or (2) the experiment has
failed and a photon has been emitted from the cavity, but the
detector has failed to detect it. There is no way of knowing for
sure that the desired entangled state has been generated. On the
other hand, in our scheme, the detection of a photon by the
detector $D_2$ assures that the desired entangled state has indeed
been generated. The finite detection efficiency only reduces the
probability for such a detection. In fact, with the automatic
feedback installed in our scheme and assuming the efficiency of
the photon feedback to be unity, we know that we have the desired
entangled state generated inside the cavity after a sufficiently
long time (after several cavity decay times), even if the detector
$D_2$ fails to click because of the finite detection efficiency.
In practical situations, however, the feedback efficiency is less
than unity, and the failure to detect the photon could be due
either to the finite efficiency of the detector or to feedback
losses. The corresponding atomic state will then be a statistical
mixture of $\ket{L,L}$ and the desired entangled state. In this
situation the experiment should be restarted from the very
beginning.

Finally, we wish to consider some practical issues in relation to
the requirements on atom trapping imposed by our scheme. Our
scheme requires that the two atoms be symmetrically coupled to the
cavity mode for the entire duration of the experiment. This means
that, as we have already assumed in Eq.(1), the coupling strength
$g_\lambda$ should be the same at all times for the two atoms. It
in turn requires that, since $g_\lambda$ depends on the position
of the atom inside the cavity, both atoms be localized within the
Lamb-Dicke limit, so that the random variation of the coupling
strength $g_\lambda$ due to thermal motion is negligible. The
Lamb-Dicke condition states that the thermal vibrational amplitude
of the atom must be small compared with the optical wavelength.
Let us assume that the atoms are trapped in the low-lying states
of the trapping potential. Let us further assume that the trapping
potential is generated from a far-off-resonance trapping (FORT)
beam\cite{12}. We take the trapping potential to be
$V(x)=-V_0\cos^2 k_Tx$ ($x$ represents the coordinate along the
cavity axis, and $k_T$ is the wave number for the FORT beam), and
approximate the potential to be a harmonic potential around the
minimum point $x=0$. The Lamb-Dicke condition can then be written
as $V_0\gg\frac{\textstyle \hbar^2k^4}{\textstyle 8mk_T^2}$ ($m$
is the mass of the cesium atom). Taking
$\lambda_T=869nm=\frac{\textstyle 2\pi}{\textstyle k_T}$ and
$\lambda=852.36nm=\frac{\textstyle 2\pi}{\textstyle k}$, this
condition yields $V_0\gg 0.5kHz$. In comparison, the value of
$V_0$ as large as $45MHz$ has been reported\cite{12}. A
challenging requirement comes from the assumption that the atoms
are trapped in the low-lying state, say the ground state, of the
trapping potential. This requires that thermal energy of the atoms
be less than the ground state energy of the trapping potential.
Taking again the case of the FORT beam, the required condition
becomes $T \le \frac{\textstyle \hbar k_T}{\textstyle
2k_B}\sqrt{\frac{\textstyle 2V_0}{\textstyle m}}$ ($k_B$ is the
Boltzman constant). Taking $\lambda_T=869nm$, and $V_0=45MHz$,
this condition yields $T\le 14\mu K$. Although the temperature as
low as $2\sim 3\mu K$ has been achieved experimentally\cite{12},
this condition on temperature pushes the present technology to its
limit. In this regard, we note that there seems to exist a
promising method, namely the adiabatic scheme recently proposed by
Duan et al.\cite{13}, which may allow a successful operation of
our experiment even with ``hot" trapped atoms beyond the
Lamb-Dicke limit. In this adiabatic scheme, by keeping the pumping
laser collinear with the cavity axis and thereby allowing the
driving pulse to have the same spatial mode structure as the
cavity mode, the system dynamics can be made to become independent
of the random atom position generated by thermal motion.

Our scheme also requires a reliable and efficient source of a
single left-circularly polarized photon. This is certainly a
difficult requirement to achieve even with much experimental
progress\cite{14} witnessed recently. One promising source that
can be used for our experiment may be a single atom trapped in a
high-Q cavity. When combined with the adiabatic scheme of Duan et
al.\cite{13}, such an atom could represent a fully controllable
single photon source.

This research was supported by the Brain Korea 21 Project of the
Korean Ministry of Education, by the Korea Research Foundation
under Contract No. DP0107, and by the Korea Science and
Engineering Foundation under Contract No. 1999-2-121-005-3. We
thank Professor D.H. Cho of Korea University for helpful
discussion.

\end{document}